\documentclass[preprint2]{aastex}

\newcommand{\kpc}{\mbox{kpc}}
\newcommand{\kpch}{\mbox{$h^{-1}$kpc}}

\newcommand{\Mpch}{\mbox{$h^{-1}$Mpc}}

\newcommand{\Msunh}{\mbox{$h^{-1}$M$_\odot$}}

\newcommand{\etal}{et al.}

\newcommand{\rs}{\mbox{$r_{\rm s}$}}
\newcommand{\rvir}{\mbox{$r_{\rm vir}$}}
\newcommand{\Mvir}{\mbox{$M_{\rm vir}$}}
\newcommand{\LCDM}{{$\Lambda$CDM}}
\newcommand{\mathnew}{\mathsurround=0pt}
\newcommand{\simov}[2]{\lower .5pt\vbox{\baselineskip0pt
    \lineskip-.5pt\ialign{$\mathnew#1\hfil##\hfil$\crcr#2\crcr\sim\crcr}}}  
\newcommand{\simgreat}{\mathrel{\mathpalette\simov >}}

\newcommand{\Cnfw}{C_{\rm NFW}}
\newcommand{\Cmoore}{C_{\rm Moore}}
\shorttitle{Resolving Dark Halos}
\shortauthors{Klypin et al.}
\begin{document}
\title{Resolving the Structure of Cold Dark Matter Halos}

\author{Anatoly Klypin}
\affil{Astronomy Department, New Mexico State University, Box 30001, Department
4500, Las Cruces, NM 88003-0001}
\author{Andrey V. Kravtsov\footnote{Hubble Fellow}, James S. Bullock}
\affil{Department of Astronomy, The Ohio State University, 140 West 18th Ave.,
Columbus, OH 43210-1173}
\author{Joel R. Primack}
\affil{Department of Physics, University of California, Santa Cruz, CA 95064 \\[3mm]
}

\begin{abstract}
  
  We present results of a convergence study in which we compare the
  density profiles of CDM dark matter halos simulated with varying mass
  and force resolution.  We show that although increasing the mass and
  force resolution allows one to probe deeper into the inner halo
  regions, the halo profiles converge at scales larger than the
  ``effective'' spatial resolution of the simulation.  This resolution
  is defined both by the force softening and by the mass resolution. On
  radii larger than the ``effective'' spatial resolution, density
  profiles do not experience any systematical trends when the number of
  particles or the force resolution increase further. In the
  simulations presented in this paper, we are able to probe density
  profile of a relaxed isolated galaxy-size halo at scales
  $r=(0.005-1)\rvir$.  We find that the density distribution at
  resolved scales can be well approximated by the profile suggested by
  Moore \etal (1998): $\rho\propto x^{-1.5}(1+x^{1.5})^{-1}$, where
  $x=r/\rs$ and \rs~ is the characteristic radius.  The analytical
  profile proposed by Navarro et al.  (1996) $\rho\propto
  x^{-1}(1+x)^{-2}$, also provides a good fit, with the same relative
  errors of about 10\% for radii larger than 1\% of the virial radius.
  For this limit both analytical profiles fit well because for
  high-concentration galaxy-size halos the differences between these
  profiles become significant only at scales well below $0.01\rvir$.
  We also find that halos of similar mass may have somewhat different
  parameters (characteristic radius, maximum rotation velocity, etc.)
  and shapes of their density profiles.  We associate this scatter in
  properties with differences in halo merger histories and the amount
  of substructure present in the analyzed halos.

\end{abstract}  

%=====================

\section{Introduction}

During the last decade there has been an increasingly growing interest
in testing  the  predictions of  variants of   cold dark  matter (CDM)
models   at subgalactic   ($\lesssim  100{\rm\  kpc}$)  scales.   This
interest was initiated by indications that observed rotation curves in
the central  regions of dark  matter  dominated dwarf galaxies are  at
odds with  predictions  of hierarchical  models.  Specifically, it was
argued (Flores  \& Primack 1994; Moore 1994) that circular velocities,
$v_c(r)\equiv[GM(<r)/r]^{1/2}$,       at small galactocentric    radii
predicted by the  models are too high   and increase too  rapidly with
increasing radius  compared   to  the observed rotation    curves. The
steeper than expected rise of $v_c(r)$ implies that the {\em shape} of
the predicted halo density  distribution is incorrect and/or  that the
DM halos  formed in  CDM models are  too concentrated  (i.e., have too
much of their mass concentrated in the inner regions).

In addition to the density profiles, there  is an alarming mismatch in
the predicted   abundance of small-mass   ($\lesssim 10^8-10^9\Msunh$)
galactic satellites and the observed number of satellites in the Local
Group (Kauffmann, White \& Guiderdoni 1993;  Klypin et al. 1999; Moore
et al. 1999).  Although  this discrepancy may  well be due to feedback
processes  (such as  photoionization)  which prevent  gas collapse and
star  formation  in the majority  of the  small-mass satellites (e.g.,
Bullock,  Kravtsov  \& Weinberg 2000),  the  mass  scale at  which the
problem sets in is similar to the  scale in the spectrum of primordial
fluctuations that  may be  responsible  for the problems  with density
profiles. In the age of precision cosmology that forthcoming {\sl MAP}
and  {\sl Planck}   cosmic microwave  background  anisotropy satellite
missions are  expected to bring,  tests of the  cosmological models at
small scales  may prove to  be  the final  frontier and the  ultimate
challenge to our understanding of cosmology and structure formation in
the Universe.   However, this obviously requires  detailed predictions
and  checks   from    the   theoretical  side,  as    well  as  higher
resolution/quality observations  and a  good  understanding  of  their
implications and  associated caveats.  In  this paper  we focus on the
theoretical predictions of the density distribution of DM halos.

A systematic study  of halo density profiles for  a wide range of halo
masses  and cosmologies  was done  by  Navarro, Frenk \& White  (1996,
1997; hereafter NFW), who argued   that the analytical profile of  the
form  $\rho(r)    =\rho_s(r/r_s)^{-1}(1+r/r_s)^{-2}$  provided a  good
description of halo profiles in  their simulations for all halo masses
and in all  cosmologies. Here, $r_s$   is the scale  radius which, for
this profile, corresponds  to  the scale at which  $d\log\rho(r)/d\log
r\vert_{r=r_s}=-2$.  The parameters of  the profile are determined  by
the  halo's virial mass  $\Mvir$  and {\em  concentration} defined  as
$c\equiv  \rvir/r_s$. NFW argued   that there is  a tight  correlation
between $c$ and $\Mvir$,  which implies that the density distributions
of   halos  of different   masses  can  in  fact    be described  by a
one-parameter family  of    analytical profiles. Further studies    by
Kravtsov, Klypin \& Khokhlov (1997), Kravtsov  et al. (1998, hereafter
KKBP98), Jing (2000), Bullock et al.   (2000), although confirming the
$c(\Mvir)$ correlation, indicated that there is significant scatter in
both the density  profiles and concentrations for  DM halos of a given
mass.

Following the  initial studies by Flores  \& Primack (1994)  and Moore
(1994),  KKBP98 presented  a  systematic comparison  of the results of
numerical simulations with  rotation curves of  a sample of  seventeen
dark matter dominated dwarf and low surface brightness (LSB) galaxies.

We pointed out that the measured rotation curves of these galaxies all
had the same shape with nearly linear central behavior, and
furthermore, based on comparison with the density profiles of
simulated halos there did not seem to be a significant discrepancy in
the {\em shape} of the density profiles at the scales probed by the
numerical simulations ($\gtrsim 0.02-0.03\rvir$, where $\rvir$ is
halo's virial radius). In other words, the central density
distribution in both galaxies and CDM halos was found to be shallower
than $\propto r^{-1}$. These conclusions were subject to several
caveats and required further testing. First, observed galactic
rotation curves had to be re-examined more carefully and with higher
resolution. The fact that all of the observed rotation curves used in
earlier analyses were obtained using relatively low resolution HI
observations required checks of the possible beam smearing effects.
Also, the possibility of non-circular random motions in the central
regions which could modify the rotation velocity of the gas (e.g.,
Binney \& Tremain 1987, p.  198) had to be considered.  Second, the
theoretical predictions had to be tested for convergence and extended
to scales $\lesssim 0.01\rvir$.

Moore et al. (1998; see also a more recent convergence study by Ghigna
et al.   1999)   presented a   convergence  study  arguing  that  mass
resolution    has  a  significant  impact     on the  central  density
distribution of halos.  They  suggested that at least several  million
particles per halo are required to reliably model the density profiles
at scales $\lesssim 0.01\rvir$.  Based on these results,  Moore et al. 
(1998) advocated  a   density profile   of the   form  $\rho(r)\propto
(r/r_s)^{-1.5}[1+(r/r_s)^{1.5}]^{-1}$,    that     behaves   similarly
($\rho\propto  r^{-3}$) to the  NFW   profile at  large radii, but  is
steeper at small $r$: $\rho\propto  r^{-1.5}$. Most recently, Jing  \&
Suto (2000) presented a systematic  study of density profiles for halo
masses  in  the range $2\times   10^{12}\Msunh-5\times 10^{14}\Msunh$. 
The   study was uniform in mass   and force resolution featuring $\sim
5-10\times  10^5$ particles per  halo and force resolution of $\approx
0.004\rvir$.  They found that  galaxy-mass  halos in their simulations
are well fitted   by  the profile\footnote{Note that  this  profile is
  somewhat different  than the profile advocated  by Moore et al., but
  behaves similarly to the latter at small radii.  Figure 9 shows that
  all three profiles --- NFW, Moore, and Jing \& Suto --- provide good
  fits to   dark   matter  halos   simulated  at  high    resolution.} 
$\rho(r)\propto (r/r_s)^{-1.5}[1+r/r_s]^{-1.5}$, but that cluster-mass
halos are well described by the NFW profile, with logarithmic slope of
the density profiles at $r=0.01\rvir$ changing from $\approx -1.5$ for
$\Mvir\sim 10^{12}\Msunh$ to  $\approx  -1.1$  for $\Mvir\sim  5\times
10^{14}\Msunh$.  Jing \& Suto interpreted these results as an evidence
that profiles of  DM halos  are not  universal (but  see  \S3.1 for  a
possible alternative interpretation).

At small scales, the results  of Kravtsov et  al.  (1998) are at  odds
with   the results   of   above studies.   Although   fairly extensive
convergence tests were done in that study, they focused on the effects
of  spatial resolution and the  mass  resolution was  kept constant in
almost all the tests. In this case we found that halo density profiles
converged at scales larger than   two  formal resolutions of the   ART
code. As we will show  in this paper, this is  not true when the  mass
resolution is varied. In  particular, the convergence  study presented
in this paper in which we varied  both mass and force resolution shows
that for  the ART simulations convergence is  reached at scales larger
than  four formal resolutions or the  scales containing 200 particles,
whichever is larger. The shallow behavior of  the density profiles in
Kravtsov et  al. was found at  scales $\approx 2-5$ formal resolutions
(at  larger scales profiles were   consistent with the NFW  functional
form) and is therefore a  numerical artifact. In simulations presented
in this paper we find profiles that are consistent  with cuspy NFW and
Moore et al. distributions at well resolved scales.

New observational  and  theoretical developments  show that comparison
between  model   predictions     and   observational data      is  not
straightforward.  Decisive comparisons require reaching convergence of
theoretical predictions and understanding the kinematics of the gas in
the central  regions of observed galaxies.  As we noted above, in this
paper  we present convergence tests  designed  to test effects of mass
resolution  on the density profiles  of  halos formed in the currently
popular   CDM model with  cosmological  constant (\LCDM) and simulated
using the multiple mass resolution version  of the Adaptive Refinement
Tree code (ART). This study is crucial in resolving the discrepancy of
our previous  study  on  the  density profiles  with  other  numerical
studies.  We  also  discuss several  caveats  with respect  to drawing
conclusions about  the density  profiles from  the fits of  analytical
functions to numerical results and  their comparisons to observational
data. In   the following section  we describe   the code and numerical
simulations  used in our analysis.  In  \S 3 we compare the analytical
fits advocated by NFW and Moore et al., fits of  these profiles to the
density  profiles of simulated  halos, and convergence analysis of our
numerical results.

%--------------------------------
\section{Numerical simulations}
%--------------------------------

\subsection{Code description}

The Adaptive Refinement  Tree code (ART;  Kravtsov, Klypin \& Khokhlov
1997) was used   to run the  simulations.  The ART code starts  with a
uniform  grid, which  covers the whole   computational box. This  grid
defines  the  lowest (zeroth) level of   resolution of the simulation. 
The standard Particles-Mesh algorithms are used to compute density and
gravitational  potential  on the   zeroth-level  mesh.  The  ART  code
reaches   high force resolution by refining   all high density regions
using   an  automated  refinement   algorithm.   The   refinements are
recursive:  the refined regions can  also be  refined, each subsequent
refinement having half of the previous level's cell size. This creates
a hierarchy of refinement  meshes  of different resolution, size,  and
geometry  covering regions of  interest. Because each individual cubic
cell can be refined, the shape of the refinement mesh can be arbitrary
and match effectively the geometry of the region of interest.

The criterion for refinement is the local density of particles: if the
number of particles in a mesh cell (as estimated by the Cloud-In-Cell
method) exceeds the level $n_{\rm thresh}$, the cell is split
(``refined'') into 8 cells of the next refinement level.  The
refinement threshold may depend on the refinement level. The code uses
the expansion parameter $a$ as the time variable.  During the
integration, spatial refinement is accompanied by temporal refinement.
Namely, each level of refinement, $l$, is integrated with its own time
step $\Delta a_l=\Delta a_0/2^l$, where $\Delta a_0$ is the global
time step of the zeroth refinement level.  This variable time stepping
is very important for accuracy of the results.  As the force
resolution increases, more steps are needed to integrate the
trajectories accurately. In the remainder of the paper by the term
{\em formal resolution}, $h_{\rm formal}$, we will mean the size of a
cell on the highest level of refinement reached in simulation. This is
similar to the usual practice of defining formal resolution in the
uniform grid codes. The actual force resolution of the code is
somewhat larger than the formal resolution. In the ART code, the
interparticle force is weaker (``softer'') than the Newtonian force at
scales $<2h_{\rm formal}$. The average interparticle force is
Newtonian at scales $\simgreat 2h_{\rm formal}$, although there is a
substantial scatter in the force at $\approx 2-3h_{\rm formal}$ due
primarily to errors in numerical differentiation of potential.  For
comparison, average Plummer softened force with softening
$\epsilon_{\rm Plummer}=h_{\rm formal}$, reaches the Newtonian value
at scales $\approx 5-6h_{\rm formal}$ (see Kravtsov et al. 1997 for
details). Therefore, one formal resolution in the ART code 
is equivalent to $\approx 0.3-0.4$ Plummer softening of the same
value. 

The cosmological simulations performed using the ART code were
compared with the simulations started from identical initial
conditions and performed using the well-known PM and AP$^3$M codes.
The comparisons showed that results of ART simulations (for a wide
battery of the commonly used statistics and halo parameters) are
similar to those of the AP$^3$M simulations at all resolved scales.
These comparisons and other tests of the ART code can be found in
Kravtsov (1999) and Knebe et al. (2000).

\subsection{Initial conditions}

The  current version   of  the ART  code  has  the  ability  to handle
particles of  different masses.  In  the present analysis this ability
was used   to  increase  the  mass  (and correspondingly  the   force)
resolution inside   a  few  pre-selected  halos.  The    multiple mass
resolution is  implemented   in the  following  way.   We   set up   a
realization of the  initial spectrum of  perturbations  in such  a way
that a very  large number of small-mass particles  can be generated in
the simulation  box.   For example,  for  the  first (second)   set of
simulations (see  below)  we generate   $512^3$ ($1024^3$) independent
spectrum  harmonics.    Potentially,     initial   conditions     with
$512^3/1024^3$  particles   could  be  generated.     Coordinates  and
velocities  of the particles  are calculated  using  all waves ranging
from   the  fundamental mode $k=2\pi/L$     to  the Nyquist  frequency
$k=2\pi/L\times N^{1/3}/2$, where $L$ is  the box size  and $N$ is the
number of particles in the simulation.

The   code actually  generates  positions    and  velocities  for  all
$512^3/1024^3$ particles, but  some of the  particles are then  merged
into particles  of larger mass. The larger   mass (merged) particle is
assigned  velocity and displacement equal  to the average velocity and
displacement of  the merged particles.  The whole lagrangian  space of
particles  is divided into large  cubic blocks  of particles with each
block  having   $16^3$ particles.     Depending  on what   local  mass
resolution is required, each  particular block can be subdivided  into
smaller   sub-blocks  and generate  from  1  to  $16^3$ particles (the
highest   resolution).  Using this   procedure\footnote{  The code  is
  actually written to handle an  arbitrary dynamic range.  The current
  limit is determined by computational  limitations.}, we can generate
particles with 5 different masses covering dynamic mass range of 4096.

\begin{figure}[tb!]
\plotone{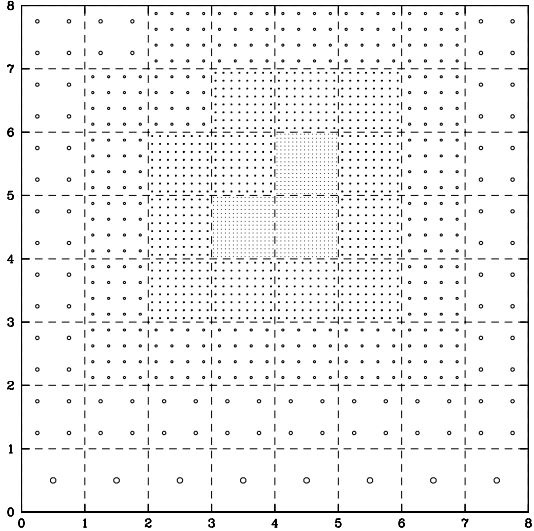}
\caption{\small Example of the construction of mass refinement in
  lagrangian space (here for illustration we show a 2D case). Three
  central blocks of particles were marked for highest mass resolution.
  Each block produces $16^2$ particles of the smallest mass. Adjacent
  blocks correspond to the four times lower resolution and produce $8^2$
  particles each. The procedure is repeated recursively until we reach
  the lowest level of resolution. The region of the highest resolution
  can have arbitrary shape.  }\label{fig:gridexample}
\end{figure}

We start simulations by making a  low resolution run with uniform mass
resolution  in  which all particles have  the  largest possible  mass. 
Next we run simulations with $32^3$  and $64^3$ particles. Using these
runs, we  identify  halos in the  simulation  and select halos   to be
re-simulated with higher mass and force resolution.  For each selected
halo  we determine its virial  radius.  We then identify all particles
inside  the two virial radii and  find lagrangian  coordinates of each
particle.   The  coordinates are used to  mark  blocks of particles to
generate  the initial conditions of the  highest mass resolution. Once
all particles are  processed and  all  blocks are marked,  we mark all
blocks adjacent to those already  marked to produce initial conditions
of the eight times lower mass resolution.   This procedure is repeated
for lower and lower mass resolution levels.  In the end, each unmarked
block will produce  one most massive particle and  a marked block will
generate a number of particles which depends  on the step in which the
block was marked.  Figure  \ref{fig:gridexample} shows the  outcome of
the process of mass refinement in a 2-dimensional case.

\begin{figure}[tb!]
\epsscale{2.05}
\plottwo{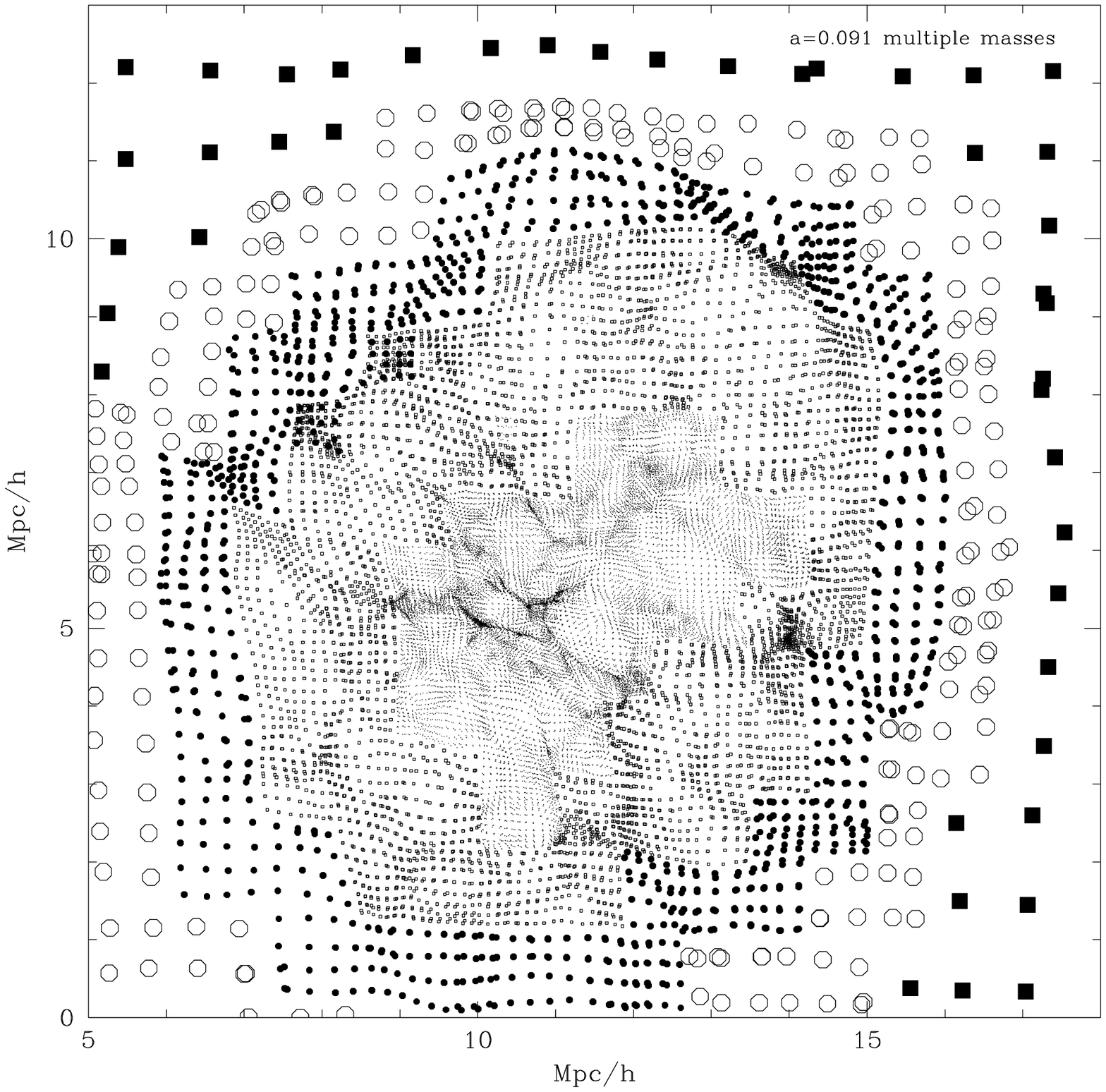}{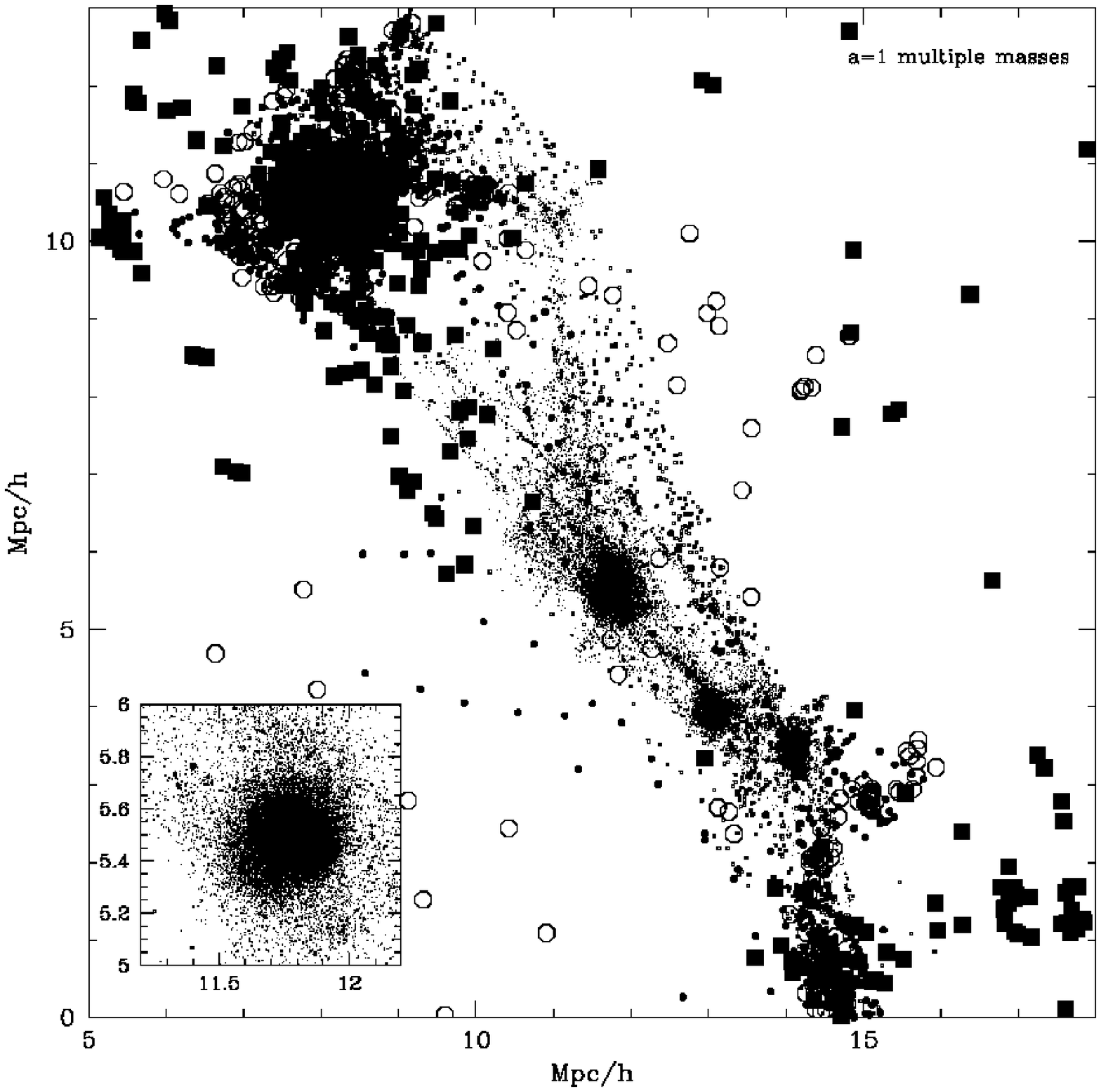}
\caption{\small Distribution of particles of different masses in a thin
slice through the center of halo A$_1$ (see Table~1) at $z=10$ (top
panel) and at $z=0$ (bottom panel). To avoid crowding of points
the thickness of the slice is made smaller in the center (about 30\kpch)
and larger (1\Mpch) in the outer parts of the forming halo. Particles
of different mass are shown with different symbols: tiny dots, 
dots, large dots, squares, and open circles.  }  
\label{fig:Massexample}
\end{figure}

Figure \ref{fig:Massexample} shows an example  of mass refinement  for
one of the  halos   in our simulations.    A large  fraction  of  high
resolution particles ends up in the  central halo, which does not have
any larger mass particles (see insert in the bottom panel). At $z=10$,
the region occupied by the high resolution particles is non-spherical:
it is substantially  elongated in the  direction perpendicular  to the
large filament clearly seen at $z=0$.
 
After   the initial conditions are set,   we  run the simulation again
allowing the code to perform mesh refinement based  only on the number
of particles with the smallest mass.

\subsection{Numerical simulations}
\label{sec:simulations}

We simulated a flat low-density cosmological model ($\Lambda$CDM) with
$\Omega_0 = 1 - \Omega_\Lambda = 0.3$, the Hubble parameter (in units
of $100{\rm\ km s^{-1} Mpc^{-1}}$) $h=0.7$, and the spectrum
normalization $\sigma_8=0.9$.  We have run two sets of simulations.
The first set used $128^3$ zeroth-level grid in a computational box of
$30\Mpch$.  The second set of simulations used $256^3$ grid in a
$25\Mpch$ box and had higher mass resolution.  In the
simulations used in this paper, the threshold for cell refinement (see
above) was low on the zeroth level: $n_{\rm thresh}(0)=2$.  Thus, every
zeroth-level cell containing two or more particles was refined.  This
was done to preserve all small-scale perturbations present in the
initial spectrum of perturbations.  The threshold was higher on deeper
levels of refinement. For the first set of simulations it was $n_{\rm
thresh}=2$ at the first refinement level and $n_{\rm thresh}=3$ for all
higher levels. For the second simulation the thresholds were $n_{\rm
thresh}=3$ and $n_{\rm thresh}=4$ for the first level and higher
levels, respectively.

In addition to effects introduced by limited mass and force
resolution, integration errors of particle trajectories may affect the
innermost regions of halos.  The local dynamical time for particles
moving in these regions is quite short.  For example, the period of a
particle on a circular orbit of radius $1\kpch$ around the center of
halo A is only $0.5\%$ of the Hubble time.  Therefore, if the time
step is not sufficiently small, numerical errors in these regions will
tend to grow. Even for small time steps  errors exist and tend to alter
the density distribution in the centers of halos over some limited 
range of scales.  

All of our simulations were started at $z_i=60$ and the step in the
expansion parameter was chosen to be $\Delta a_0=2\times 10^{-3}$ for
particles located on the zeroth base grid.  This gives about 500 steps
for particles located in the zeroth level for an entire run to $z=0$.
We have done a test run with twice smaller time step for a halo of
mass comparable (but with smaller number of particles) to the mass of
halos studied in this paper. We did not find any significant
differences in the resulting halo profile.  For both sets of
simulations, the highest level of refinement was ten for the largest
mass resolution, which corresponds to $500\times 2^{10}\approx
500,000$ time steps at the tenth refinement level. Some simulations
were rerun with smaller number of particles. They did not reach the
highest levels of refinement, and, thus, they had fewer steps. For
example, halo D$_2$ has reached only 7 levels of refinement and had
only $500\times 2^{7}\approx 64,000$ time steps.

In the following  sections we present density  profiles of four halos. 
The halo A was the only halo selected for re-simulation in the first
set of  simulations. It was relatively quiescent  at  $z=0$ and had no
massive neighbors. The halo was located in a long filament bordering a
large void  and was about  10~Mpc  away from  the nearest cluster-size
halo.  After the high-resolution  simulation  was completed we   found
that the nearest galaxy-size halo was about 5~Mpc  away.  The halo had
a fairly typical merging history with $M(t)$ track slightly lower than
the average mass  growth predicted using the  extended Press-Schechter
model.  The last major  merger event  occurred  at $z\approx 2.5$;  at
lower redshifts the  mass growth (the mass  in this  time interval has
grown by a factor of three) was due to slow and steady mass accretion.

The halos B, C, and D were identified in the second set of
simulations and were selected among halos residing in a well defined
filament.  Two of the halos (B and C) are neighbors located about
0.5~Mpc from each other. The third halo was 2~Mpc away from this pair.
Thus, the halos were not selected to be too isolated as was the case
in the first set of runs. Moreover, the simulation was also analyzed
at both $z=0$ and at $z=1$ (when halos are more likely to be less
relaxed).  Therefore, the halo A can be considered as an example
of a rather isolated well-relaxed halo. In many respects, this halo is
similar to halos simulated by other research groups that used multiple
mass resolution techniques.  The halos B, C, and D from the
second set of simulations can be viewed as representative of more
typical halo population located in more crowded environments.

Parameters of the simulated dark matter halos are listed in Table~1.
Columns in the table present (1) halo's ``name'' (halos A$_1$, A$_2$,
A$_3$ are the halo A re-simulated three times with different mass and
force resolutions); (2) redshift at which the halo was analyzed; (3)
the number of particles within the virial radius; (4) the smallest
particle mass in the simulation; (5) formal comoving force resolution
(cells size at the highest refinement level) achieved in the
simulation.

\begin{deluxetable}{ccccc} 
\tablecolumns{11} 
\tablewidth{0pc} 
\tablecaption{Simulation parameters} 
\tablehead{ 
\colhead{\small Halo} & \colhead{z}  & \colhead{$N_{\rm part}$}   &
\colhead{$m_{\rm part}$}    & \colhead{\small h$_{\rm formal}$} \\
\colhead{} & \colhead{} & \colhead{} & \colhead{$h^{-1}M_{\odot}$} 
& \colhead{$h^{-1}$kpc} \\
\colhead{(1)} &\colhead{(2)} &\colhead{(3)} &\colhead{(4)} &\colhead{(5)} 
}
\startdata 
A$_1$ & 0 &$1.2\times 10^5$ & $1.6\times 10^7$ & 0.23 \\ 
A$_2$ & 0 &$1.5\times 10^4$ & $1.3\times 10^8$ & 0.91 \\ 
A$_3$ & 0 &$1.9\times 10^3$ & $1.1\times 10^9$ & 3.66 \\ 
B$_1$ & 0 &$1.0\times 10^6$ & $1.2\times 10^6$ & 0.10 \\ 
B$_2$ & 0 &$1.5\times 10^4$ & $1.2\times 10^6$ & 0.76 \\ 
B$_3$ & 1 &$7.1\times 10^5$ & $1.2\times 10^6$ & 0.19 \\ 
C$_1$ & 0 &$1.1\times 10^6$ & $1.2\times 10^6$ & 0.10 \\ 
C$_2$ & 0 &$1.6\times 10^4$ & $7.7\times 10^7$ & 0.76 \\ 
C$_3$ & 1 &$5.0\times 10^5$ & $1.2\times 10^6$ & 0.19 \\ 
D$_1$ & 0 &$1.3\times 10^6$ & $1.2\times 10^6$ & 0.10 \\ 
D$_2$ & 0 &$2.0\times 10^4$ & $7.7\times 10^7$ & 0.76 \\ 
D$_3$ & 1 &$7.9\times 10^5$ & $1.2\times 10^6$ & 0.19 \\ 
\enddata 
\end{deluxetable} 

%-----------------
\section{Results}
%-----------------

\subsection{Comparison of the NFW and the Moore et al. profiles}

Before we fit analytical profiles to profiles of simulated dark matter
halos  or  compare   them to the   observed   rotation curves,  it  is
instructive  to compare  different analytical approximations. Although
the NFW   and Moore  et  al. profiles   predict different behavior  of
$\rho(r)$ in the central  regions of a halo,  the scale at  which this
difference becomes significant depends  on the specific values of  the
halo's  characteristic  density  and  radius.   Table~2   presents the
parameters and statistics associated with the two analytical profiles.
For  the NFW profile  more information can be found  in  Klypin et al. (1998), 
 \L okas \& Mamon (2000), and Widrow (2000).
 
 Each  profile is  defined by two  independent  parameters.  We choose
 these to be the characteristic density  $\rho_s$ and radius $r_s$. In
 this case all expressions  describing the properties of the  profiles
 have a simple form and do not depend  on the concentration.  Both the
 concentration and the virial mass appear only in the normalization of
 the expressions.  The choice  of the virial  radius (e.g., \L okas \&
 Mamon  2000) as a scale  unit results in more complicated expressions
 with explicit dependence on the concentration.  In this case, one has
 to be careful about the definition of the virial radius, as there are
 several  definitions  in the  literature.  For example,   it is often
 defined as the radius, $r_{200}$, within which the average density is
 200 times the {\em critical density}. In this paper the virial radius
 is defined as the radius within which the average density is equal to
 the density predicted by the  top-hat model: it is $\delta_{\rm  TH}$
 times the {\em average matter density} in  the Universe.  In the case
 of  $\Omega_0=0.3$  models the virial  radius  defined in this way is
 about 30\% larger than $r_{200}$ (e.g., Eke et al. 1998).

\begin{figure}[tb!]
\epsscale{1.0}
\plotone{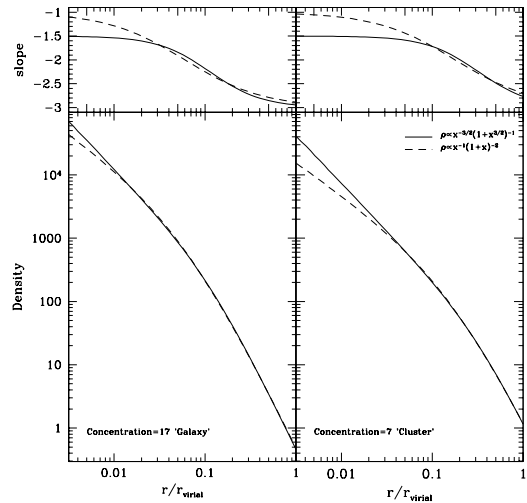}
\caption{\small Comparison of the Moore et al. and the NFW profiles. 
  Each profile is normalized to have the same virial mass and the same
  radius of the maximum circular velocity.  {\it Left panels:} 
High-concentration halo typical of small galaxy-size halos
  $C_{\rm NFW}=17$. {\it Right panels:} Low-concentration halo 
  typical of cluster-size halos. The deviations are
  very small ($<3\%$) for radii $r>r_s/2$. Top panels show the local
  logarithmic slope of the profiles. Note that for the high
  concentration halo the slope of the profile is significantly larger
  than the asymptotic value -1 even at very small radii $r \approx
  0.01r_{\rm vir}$.  }\label{fig:CompareBenNFW}
\end{figure}

There is no unique way of  defining a consistent concentration for the
different   analytical profiles. Again,   it   is natural to  use  the
characteristic radius  $r_s$   to define the  concentration:  $c\equiv
r_{\rm vir}/r_s$.  This simplifies the expressions.  At the same time,
if we fit  the dark matter  halo  with the  two profiles,  we will get
different concentrations because the values of the corresponding $r_s$
will  be different.  Alternatively,  if we choose  to  match the outer
regions of the profiles (say, $r> r_s$) as closely as possible, we may
choose to  change the  ratio of the  characteristic radii  $r_{s,  \rm
  NFW}/r_{s, \rm  Moore}$ in such a  way that both  profiles reach the
maximum circular velocity  $v_{\rm circ}$ at  the same physical radius
$r_{\rm max}$. In this case, the formal concentration  of the Moore et
al. profile  is  1.72 times smaller   than that of  the  NFW  profile. 
Indeed, with this   normalization profiles look  very similar  in  the
outer parts as one  finds in Figure  \ref{fig:CompareBenNFW}.  Table~2
also gives two  other ``concentrations''. The concentration  $C_{1/5}$
is defined  as the   ratio of  virial radius   to  the  radius,  which
encompasses 1/5  of the virial   mass (Avila-Reese et al.  1999).  For
halos  with $C_{\rm NFW}\approx   5.5$ this 1/5  mass concentration is
equal to $C_{\rm  NFW}$. One can  also define the concentration as the
ratio  of the virial  radius to  the  radius at which the  logarithmic
slope of the density profile is equal  to $-2$. This scale corresponds
to $r_s$ for  the NFW profile and $\approx  0.35 r_s$ for the Moore et
al.  profile.

\begin{deluxetable}{l|l|l} 
\tablecolumns{3} 
\tablewidth{0pc} 
\tablecaption{Comparison of  NFW and Moore et al. profiles} 
\tablehead{ 
\colhead{Parameter} & \colhead{NFW}   & \colhead{Moore et al.}  
} 
\startdata 
Density &  $\rho = \frac{\displaystyle \rho_s}{\displaystyle x(1+x)^2}$ & 
 $\rho = \frac{\displaystyle \rho_s}{\displaystyle x^{1.5}(1+x)^{1.5}}$ \\ 
\quad $x=r/r_s$ & $\quad \rho \propto x^{-3}$ for $x\gg 1$ & $\quad \rho \propto x^{-3}$ for $x\gg 1$ \\
                        & $\quad \rho \propto x^{-1}$ for $x\ll 1$ & $\quad \rho \propto x^{-1.5}$ for $x\ll 1$ \\
                 & $\quad \rho/\rho_s =1/4\phm{.00}$ at $x=1$ &  $\quad \rho/\rho_s =1/2\phm{.00}$ at $x=1$ \\
\tableline
Mass     & & \\
\quad $M=4\pi\rho_sr_s^3f(x)$ & 
 $f(x)= \ln(1+x) -\frac{\displaystyle x}{\displaystyle 1+x}$ & $f(x)= \frac{2}{3}\ln(1+x^{3/2})$ \\
\quad   $\phm{M}=M_{\rm vir}f(x)/f(C)$ & & \\
\quad  $M_{\rm vir}=\frac{4\pi}{3}\rho_{\rm cr}\Omega_0\delta_{\rm top-hat}r_{\rm vir}^3$ & \\
\tableline
Concentration &$\Cnfw = 1.72\Cmoore$   & $\Cmoore = \Cnfw/1.72$   \\
 & \quad {\small for halos with the same $M_{\rm vir}$ and $r_{\rm max}$}&  \\
\quad $C=r_{\rm vir}/r_s$ & $C_{1/5} \approx \frac{\displaystyle \Cnfw}{\displaystyle 0.86f(\Cnfw)+0.1363}$ &
                                              $C_{1/5} = \frac{\displaystyle \Cmoore}{[\displaystyle (1+\Cmoore^{3/2})^{1/5}-1]^{2/3}} $ \\
                                        &\phm{..}{\small error less than 3\% for $\Cnfw=$5-30} & \phm{C0.0}$
                                                     \approx \frac{\displaystyle \Cmoore}{\displaystyle [\Cmoore^{3/10}-1]^{2/3}}$ \\
      & $C_{\gamma=-2}=\Cnfw$ & $C_{\gamma=-2}=2^{3/2}\Cmoore$ \\
      &  & \phm{$C_{\gamma=-2}=$}$\approx 2.83\Cmoore$ \\
\tableline
Circular Velocity &  & \\
\quad $v_{\rm circ}^2 =\displaystyle \frac{GM_{\rm vir}}{r_{\rm vir}}\frac{C}{x}\frac{f(x)}{f(C)}$ &
             $x_{\rm max} \approx 2.15$ &  $x_{\rm max} \approx 1.25$ \\
\quad  $ \phm{v_{\rm circ}^2}=\displaystyle v_{\rm max}^2\frac{x_{\rm max}}{x}\frac{f(x)}{f(x_{\rm max})} $ &
         $v_{\rm max}^2 \approx 0.216v_{\rm vir}^2\displaystyle \frac{C}{f(C)}$&
          $v_{\rm max}^2 \approx 0.466v_{\rm vir}^2\displaystyle \frac{C}{f(C)}$ \\
\quad $v_{\rm vir}^2\phd=\displaystyle \frac{ GM_{\rm vir}}{r_{\rm
vir}}$         &  $\rho/\rho_s \approx 1/21.3$ at $x=2.15$
  &  $\rho/\rho_s \approx 1/3.35$ at $x=1.25$ 
\\
\enddata 
\end{deluxetable} 

Figure  \ref{fig:CompareBenNFW}  presents the  comparison between  the
analytic profiles normalized to have the same virial mass and the same
radius $r_{\rm  max}$.  We  show results  for halos   of low and  high
values of concentration representative of cluster- and low-mass galaxy
halos,  respectively.  The bottom  panels show the profiles, while the
top panels  show the corresponding  logarithmic slope as a function of
radius.  The figure  shows that  the two   profiles are very   similar
throughout the main body of the halos. Only in the very central region
do the differences become significant. The difference is more apparent
in the  logarithmic   slope than  in  the   actual  density profiles.  
Moreover, for galaxy-mass  halos the  difference  sets in  at a rather
small radius $\lesssim 0.01\rvir$, which would correspond to scales $<
1{\rm\ kpc}$ for  the typical  dark   matter dominated dwarf and   LSB
galaxies.  At the  observationally interesting scales the  differences
between NFW and  Moore et al.  profiles are  fairly small  and the NFW
profile provides   an  accurate  description  of the    halo   density
distribution.
 
Note also  that for galaxy-size  (e.g.,  high-concentration) halos the
logarithmic slope of  the NFW profile has not yet reached its  asymptotic
inner   value  of  $-1$ even  at  scales  as  small   as   $0.01\rvir$.  
%%Figure~\ref{fig:am} shows the logarithmic  slope of the NFW profile at
%%scales $\approx 0.01-0.02\rvir$ as a function  of halo virial mass for
%%the {\LCDM} model studied here. The average $c_{\rm vir}(M)$ for this
%%model given in Bullock  et al. (2000) is used here.  
At this distance  the logarithmic slope of the  NFW profile is $\approx -1.4-1.5$
for halos with mass   $\sim 10^{12}\Msunh$.
For cluster-size halos    this slope is  $\approx  -1.2$. This
dependence of the  slope at a given fraction  of the  virial radius on
the virial mass of the halo is very similar  to the results plotted in
Figure~3  of  Jing \& Suto   (2000).  These authors interpreted  it as
evidence that halo profiles are not universal. It is obvious, however,
that their results are consistent with NFW profiles and the dependence
of the slope on mass can be simply a manifestation of the well-studied
$c_{\rm vir}(M)$ relation.

%%In Figure \ref{fig:BenNFW}  we compare   the NFW and    Moore et al.   
%%profiles in a different way. 
The NFW and    Moore et al.  profiles can be compared  in a different way. 
 We can approximate  the Moore et al. halo of
a given concentration with the  NFW profile. 
Fractional deviations  of the fits depend on the halo concentration
and on the range of radii used for the  fits.
%%are shown in the figure for  halos of different concentration
%%and for three ranges of  radii used for the  fits: $0.003, 0.01, 0.02<
%%r/\rvir < 1$.  
A low-concentration halo has larger deviations, but even for $C=7$
case, the deviations are less than 15\% if we fit the halo at scales
$0.01< r/\rvir < 1$.  For a high-concentration halo with $C=17$, the deviations
are much smaller: less than 8\% for the same range of scales.

To summarize, we find that  the  differences between  the NFW and  the
Moore  et al. profiles  are very small  ($\Delta\rho/\rho < 10\%$) for
radii  above 1\% of the virial  radius for typical galaxy-size halos
with $C_{\rm NFW}\simgreat 12$. The differences are larger for
halos with smaller concentrations. In the case of the NFW profile, the
asymptotic value  of the central  slope  $\gamma =-1$ is  not achieved
even at radii as small as 1\%-2\% of the virial radius.

%%\begin{figure}[tb!]
%%\epsscale{1.0}
%%\plotone{am.eps}
%%\caption{\small Logarithmic slope at the scale $0.01-0.02\rvir$ for the 
%%NFW profile as a function of halo virial mass. The {\LCDM} cosmology and 
%%$C_{\rm vir}(M)$ appropriate for this model at $z=0$ (see Bullock et al. 2000) 
%%are assumed to compute $\gamma_{0.01}$. 
%%}\label{fig:am}
%%\end{figure}

%%\begin{figure}[tb!]
%%\epsscale{1.0}
%%\plotone{fitBenNFW17.ps}
%%\caption{\small Density deviations of the halos profiles. A halo 
%%with the  Moore et al. profile was approximated by the NFW
%%profile. Different ranges of radii were used to make the fit as
%%indicated in the plot.  
%%{\it Left panel:} High concentration $C_{\rm NFW}=17$ halo. {\it Right panel:} Low
%%concentration  $C_{\rm NFW}=7$ halo.}\label{fig:BenNFW}
%%\end{figure}

%------------------------------
\subsection{Convergence study}
%------------------------------

The effects of numerical resolution can be studied by resimulating the
same objects with higher force  and mass resolution  and with a larger
number of time  steps. In this  study we performed simulations  of the
same halos  with increasingly higher mass resolution.  In the ART code
simulations the  subsequent  mesh refinements are done   when particle
density in   a mesh  cell  exceeds a  specified   threshold.  The mass
resolution is thus tightly linked with  the highest achievable spatial
resolution. 

\begin{figure}[tb!]
\epsscale{1.5}
\plotone{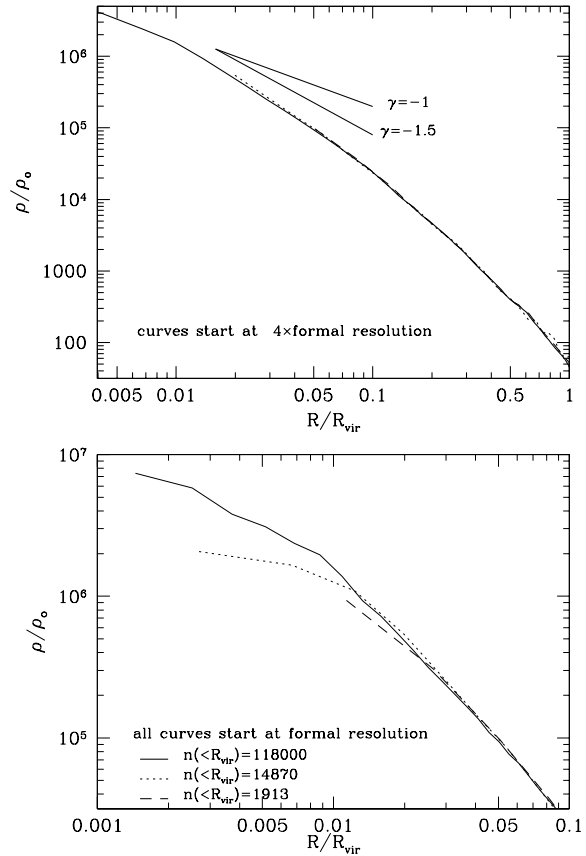}
\caption{\small Density profiles of halo  A simulated with different
mass and force resolutions. {\it Bottom panel:} The profiles are
plotted down to the formal force resolution of each simulation.
Because the plot shows results below the actual force resolution, one
gets a wrong impression that the profile gets steeper and the
concentration increases when the mass resolution is increased.  {\it
Top panel:} the profiles plotted down to {\em four} formal resolutions.
For vastly different mass and force resolutions the convergence is
reached at these scales.  At the scale of a few percent of the virial
radius the density profile is visibly steeper than the limiting slope
$\gamma=-1$ of the NFW profile. This is consistent with the NFW profile
for a halo of this concentration.
}\label{fig:Benstyle}
\end{figure}

\begin{figure}[tb!]
\epsscale{0.95}
\plotone{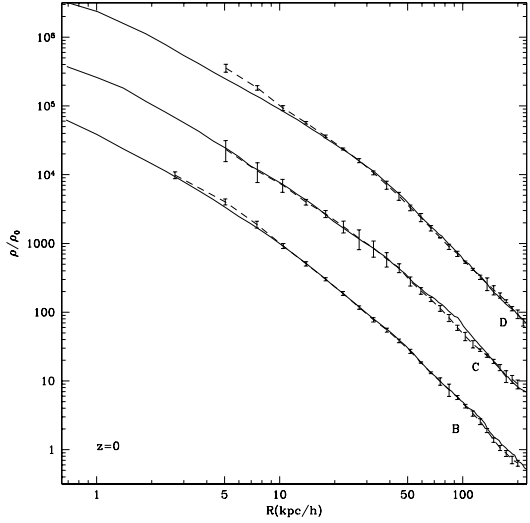}
\caption{\small Convergence of profiles for  halos B, C, D. All the curves in this plot
  start at radii, which contain at least 200 particles and which are
  larger than 4 formal resolutions. For clarity, the profiles of halos
  C and B were shifted downwards by factors of 10 and 100,
  respectively. The full curves are for halos B$_1$, C$_1$, D$_1$,
  which have more than a million particles each. The dashed curves
  show profiles of halos B$_2$, C$_2$, D$_2$ with $(15-20)\times 10^3$
  particles. The error bars indicate $2\sigma$ fluctuations of density
  at each radius due to moving satellites and residual oscillations
  inside the halos. The dispersion was estimated using density
  profiles at four time moments in the interval $z=0-0.03$ ($\approx
  4\times 10^8$~yrs).  }\label{fig:Converge}
\end{figure}

Table 3 gives parameters of the halos and parameters of their fits. The first
 and the second columns give the halo name (Table 1) and the
redshift at which the halo was studied. Columns (3-5) present virial
mass, radius, and the maximum circular velocity of the halo. Columns
(6-8) present parameters of the fits: the halo concentration as estimated
using the NFW profile and the maximum relative errors of the NFW and
Moore et al fits.  
The bottom panel in figure~\ref{fig:Benstyle} shows density profiles
for the simulations of halo A (see Table~1). Here, as in the Fig.1a in
Moore et al.  (1998), all profiles are plotted down to the formal
force resolution of the corresponding run.  Although it may appear
that density profiles have not converged in the central region and
that the low resolution simulations produce erroneous results, this is
simply an artifact of plotting the profiles below the actual numerical
resolution (or convergence scale).  The top panel in
figure~\ref{fig:Benstyle} and figure~\ref{fig:Converge} show profiles
of halos A, B, C, and D plotted down to four formal resolutions of the
simulations (4 mesh cells at the highest refinement level). The
figures show that in this case density profiles in lower resolution
simulations are in very good agreement with profiles in
high-resolution runs at all radii. For example, there are no
systematic differences in the logarithmic slope of the profile at a
given distance and we find no significant change in the concentration
parameters or  the maximum circular velocities of halos (see Table 3).

\begin{deluxetable}{cccccccc} 
\tablecolumns{11} 
\tablewidth{0pc} 
\tablecaption{Halo parameters} 
\tablehead{ 
\colhead{\small Halo} & \colhead{z}   & \colhead{$M_{\rm vir}$}    & \colhead{$R_{\rm vir}$} & \colhead{$V_{\rm max}$} 
& \colhead{$C_{\rm NFW}$}    & 
\colhead{\small RelErr}  & \colhead{\small RelErr} \\
\colhead{} & \colhead{}   & \colhead{$h^{-1}M_{\odot}$}    & \colhead{$h^{-1}$kpc} & 
\colhead{km/s} & \colhead{} &
\colhead{\small NFW} & \colhead{\small Moore} \\
\colhead{(1)} &\colhead{(2)} &\colhead{(3)} &\colhead{(4)} &\colhead{(5)} &\colhead{(6)}
&\colhead{(7)} &\colhead{(8)} }
\startdata 
A$_1$ & 0 & $2.0\times 10^{12}$ & 257 &247.0 & 17.4& 0.17& 0.20 \\ 
A$_2$ & 0 & $2.1\times 10^{12}$ & 261 &248.5 & 16.0& 0.13& 0.16 \\ 
A$_3$ & 0 & $2.0\times 10^{12}$ & 256 &250.5 & 16.6& 0.16& 0.10 \\ 
B$_1$ & 0 & $1.2\times 10^{12}$ & 215 &199.5 & 15.6& 0.30& 0.14 \\ 
B$_2$ & 0 & $1.1\times 10^{12}$ & 213 &205.0 & 16.5& 0.15& 0.14 \\ 
B$_3$ & 1 & $8.5\times 10^{11}$ & 241 &195.4 & 12.3& 0.23& 0.16 \\ 
C$_1$ & 0 & $1.3\times 10^{12}$ & 225 &190.6 & 11.2& 0.29& 0.23 \\ 
C$_2$ & 0 & $1.2\times 10^{12}$ & 220 &184.9 & 9.8 & 0.11& 0.12 \\ 
C$_3$ & 1 & $6.8\times 10^{11}$ & 208 &165.7 & 11.9& 0.37& 0.20 \\ 
D$_1$ & 0 & $1.5\times 10^{12}$ & 235 &213.9 & 11.9&0.15 & 0.68 \\ 
D$_2$ & 0 & $1.5\times 10^{12}$ & 234 &216.8 & 13.4&0.10 & 0.09 \\ 
D$_3$ & 1 & $9.6\times 10^{11}$ & 245 &202.4 &  9.5&0.25 & 0.60 \\ 
\enddata 
\end{deluxetable}

Closer  examination of the  bottom panel in Figure~\ref{fig:Benstyle},
shows that profiles have not converged at two formal resolutions. This
is at  odds with our convergence study  in Kravtsov  et al. (1998). We
attribute this  difference to the  fact  that mass  resolution in the
latter study  was kept  fixed when  force resolution was  varied.  Our
highest resolution run   A$_1$, if considered including  scales larger
than two formal resolutions, is  consistent with conclusion about  the
shallow central slope made in Kravtsov et al.  (1998).  Indeed, if the
profiles  are considered  down to    the  scale of {\em two}    formal
resolutions  (a scale smaller than  the smallest converged scale), the
density  profile slope  in the   very   central part  of the   profile
$r\lesssim 0.01r_{\rm  vir}$ is close to  $\gamma=-0.5$.   In light of
the results shown in Figure~\ref{fig:Benstyle},  it is clear that this
is an artifact     of  underestimating true  convergence    scale  and
conclusions about   shallow  central density   distributions made   in
Kravtsov et al. (1998) are therefore incorrect.

It is not clear which numerical effect determine the convergence
scale.  It is likely that this scale is determined by a complex
interplay of all numerical effects. For the ART simulations we found
empirically that the scale above which the density does not deviate
(deviations were less than 10\%) from results of higher resolution
simulation is $\gtrsim 4$ formal force resolutions or containing more
than 200 particles, whichever is larger. The limit very likely depends
on particular code used and is not universal.
Figure~\ref{fig:Benstyle}, top panel, shows that for the halo A,
convergence for vastly different mass and force resolution is reached
for scales $\gtrsim 4$ formal force resolutions (all profiles in this
figure are plotted down to the radius of 4 formal force resolutions).
For all resolutions, there are more than 200 particles within the
radius of four resolutions from the halo center.  For the highest
resolution simulation (halo A$_1$) convergence is reached at scales
$\gtrsim 0.005\rvir$, assuming convergence at 4 times the formal
resolution as found for halos A$_2$ and A$_3$.  For halos B, C, D
(figure~\ref{fig:Converge}) this criterion also worked, but was mostly
defined by the number of particles (more than 200-300 particles for
convergence).

%\begin{figure}[tb!]
%\epsscale{1.0}
%\plotone{profile0_ben1.ps}
%\caption{\small  The same as in Figure~\ref{fig:Benstyle} but with the profiles
%  plotted down to {\em four} formal resolutions.  It is clear that for
%  vastly  different mass (from 2000 to   120000 particles in the halo)
%  and   force   (from  $3.66\kpch$ to   $0.23\kpch$)   resolutions the
%  convergence  is reached  at these scales.   At  the  scale of a  few
%  percent of the virial radius  the density profile is visibly steeper
%  than the limiting slope   $\gamma=-1$ of the NFW   profile. However,
%  this is due to the high concentration of  the halo and should not be
%  interpreted as a failure of the NFW profile.  }
%\label{fig:Converge}
%\end{figure}

%==================================
\subsection{Halo profiles}
%==================================

\begin{figure}[tb!]
\epsscale{1.2}
\plotone{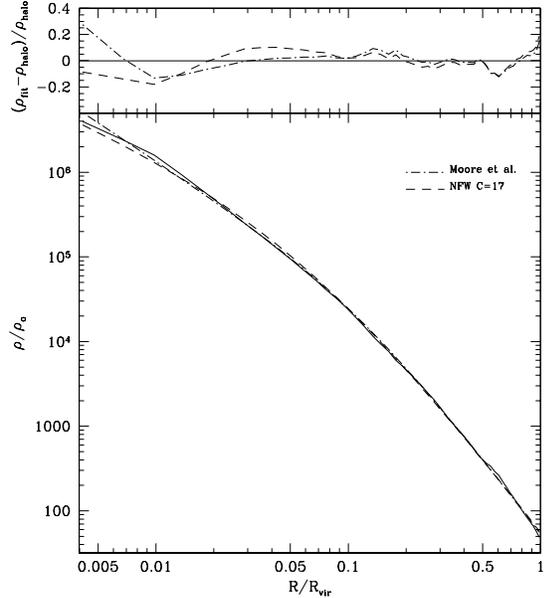}
\caption{\small  Fits of the NFW and Moore et al. halo profiles
  to the profile of halo A$_1$ ({\em bottom panel}). The {\em top
    panel} shows fractional deviations of the analytic fits from the
  numerical profile.  Note that both analytic profiles fit numerical
  profile equally well: fractional deviations are smaller than 20\%
  over almost three decades in radius. } \label{fig:FitsConverge}
\end{figure}

\begin{figure*}[tb!]
\includegraphics[width=\textwidth]{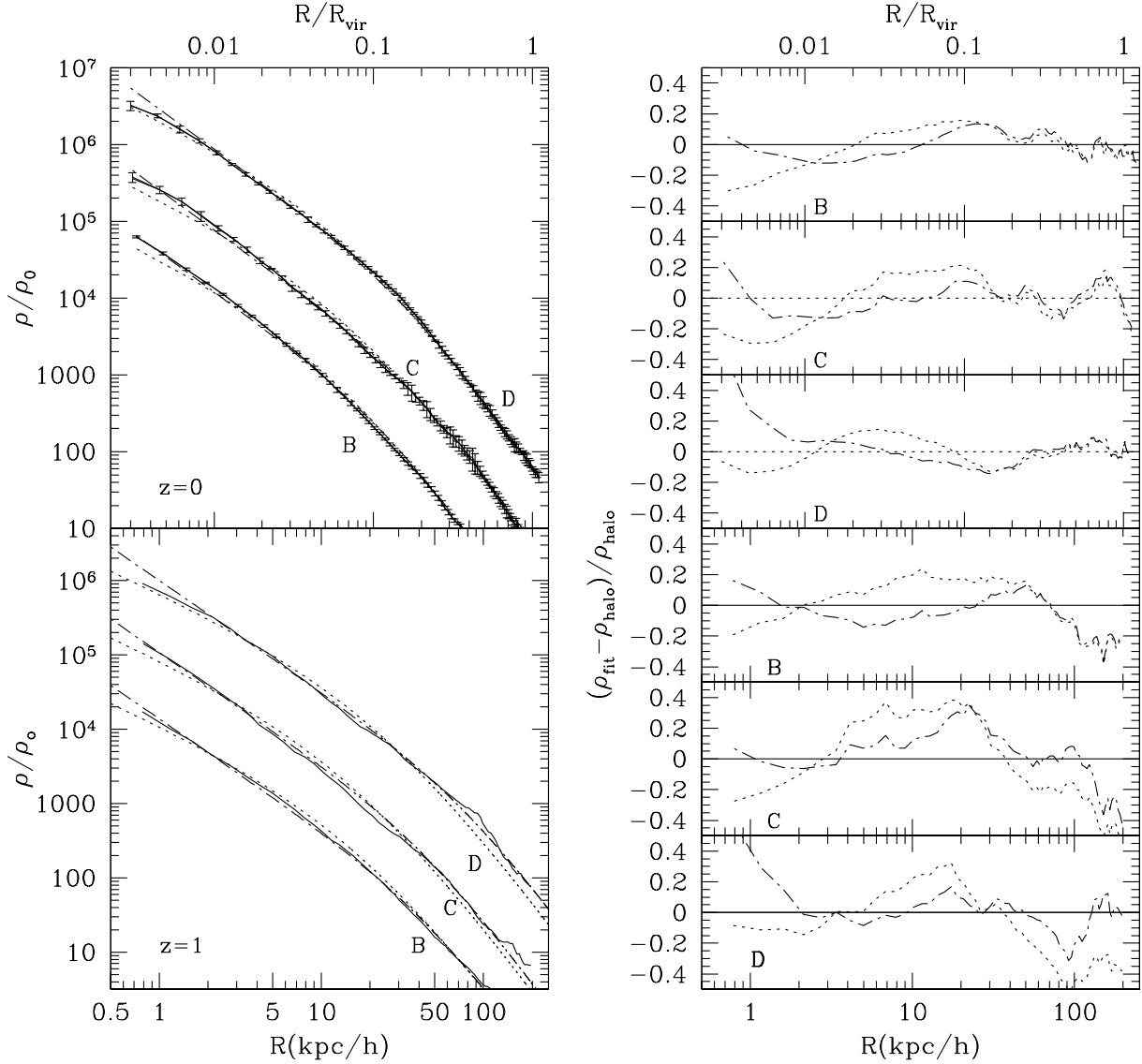}
\caption{\small  
  Fits of the NFW (dotted curves) and Moore et al. (dot-dashed curves)
  profiles to the density distributions of halos B, C, D (solid
  curves) in the highest resolution run at $z=0$ ({\em top left}).
  For clarity, the profiles of halos C and B were shifted down by
  factors of 10 and 100, respectively. Error bars present $2\sigma$
  fluctuations of density profiles at different time moments estimated
  over the last $\approx 4\times 10^8$yr of the run. {\em Top right}
  panel shows fractional deviations of the fitted profile from the
  simulated halo profile as a function of scale. {\em The bottom
    panels} show the same but for the halos B, C, and D at $z=1$. Note
  that halos at this redshift are not yet well relaxed as indicated by
  the density fluctuations in the peripheral parts of the profiles.
  } \label{fig:Profz}
\end{figure*}

In order to judge which analytical profile provides a better
description of the simulated profiles we fitted the NFW and Moore et
al. analytic profiles.  Figure~\ref{fig:FitsConverge} presents results
of the fits for halo A and shows that both profiles fit the simulated
profile equally well: fractional deviations of the fitted profiles from
the numerical one are smaller than 20\% over almost three decades in
radius. It is thus clear that the fact that the numerical profile has
slope steeper than $-1$ at the scale of $\sim 0.01\rvir$ does not mean
that a good fit of the NFW profile (or even analytic profiles with
shallower asymptotic slopes) cannot be
obtained. Figure~\ref{fig:Profz} shows the fitting of halos
in the second set of simulations. Each halo in the plot has more than a
million particles -- ten times more than halo A. One would naively
expect that this increase in the resolution should clearly show which
profile makes a better fit. Indeed, more particles and better
resolution gave smaller deviations, but the fits became better for {\it both}
approximations. For example, at 1\% of the virial radius of the halo D the deviations
were 3.6\% for the NFW profile and 6.2\% for the Moore et al. profile --
down from 20\% for the halo A at the same distance. The Moore et
al. approximation gave a better fit for halos B and C, but not for halo
D. The NFW approximation was less accurate on intermediate scales
around $(0.03-0.1)R_{\rm vir}$, but the errors were quite small. Thus,
both approximations gave comparable results. 

\begin{figure}[tb!]
\epsscale{1.1}
\plotone{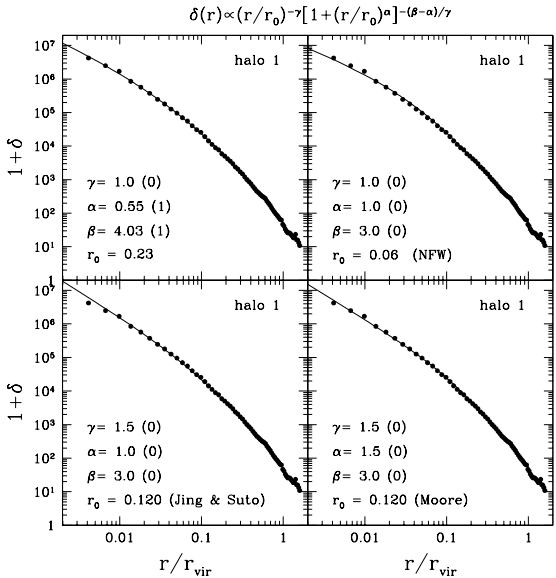}
\caption{\small Analytic fits to the density profile of the halo A$_1$ 
   from our set of simulations. The fits are 
  of the form $\rho(r)\propto (r/r_0)^{-\gamma}
  [1+(r/r_0)^{\alpha}]^{-(\beta-\alpha)/\gamma}$.  The legend in each
  panel indicates the corresponding values of $\alpha$, $\beta$, and
  $\gamma$ of the fit; the digit in parenthesis indicates whether the
  parameter was kept fixed ($0$) or not ($1$) during the fit. Note
  that various sets of parameters $\alpha$, $\beta$, $\gamma$ provide
  equally good fits to the simulated halo profile in the whole range
  resolved range of scales $\approx 0.005-1\rvir$. This indicates 
  a large degree of degeneracy in parameters $\alpha$, $\beta$, and
  $\gamma$} \label{fig:rf4_1}
\end{figure}

There is definitely a certain degree of degeneracy in fitting various
analytic profiles to numerical results.  Figure~\ref{fig:rf4_1}
illustrates this further by showing results of fitting profiles (solid
lines) of the form $\rho(r)\propto
(r/r_0)^{-\gamma}[1+(r/r_0)^{\alpha}]^{-(\beta-\alpha)/\gamma}$ to {\em
the same} simulated halo profile (halo A$_1$) shown as solid
circles. The legend in each panel indicates the corresponding values of
$\alpha$, $\beta$, and $\gamma$ of the fit; the digit in parenthesis
indicates whether the parameter was kept fixed ($0$) or not ($1$)
during the fit.  The two right panels show fits of the NFW and Moore et
al.  profile; the bottom left panel shows fit of the profiles used by
Jing \& Suto (2000). The top left panel shows a fit in which the inner
slope was fixed but $\alpha$ and $\beta$ were fit.  The figure shows
that all four analytic profiles can provide a good fit to the numerical
profile in the whole range of resolved scales: $0.005-1\rvir$.

As we mentioned in \S~\ref{sec:simulations}, the halo A analyzed in
the previous section is somewhat special because it was selected as an
isolated relaxed halo. Halos, which are not very isolated and relaxed
are also interesting. After all, they represent the majority of all
halos. When we compare observed rotation curves of galaxies with
predicted circular velocity curves of the halos, we do not know if the
galaxy host halo is well relaxed or not.  In order to reach unbiased
conclusions, we will present analysis of halos from the second set of
simulations at redshifts $z=0$ and $z=1$ (halos B$_{1,3}$, C$_{1,3}$,
and D$_{1,3}$), which were not selected to be relaxed or isolated.
Note that these halos did not have major mergers immediately prior to
the epoch of analysis, which could produce large distortions of their
profiles.  Based on the results of the convergence study presented in
the previous section, we will consider profiles of these halos only at
scales above four formal resolutions and not less than 200 particles.
There is an advantage in analyzing halos at a relatively high
redshift.  Halos of a given mass will have lower concentration (see
Bullock et al.  2000). Lower concentration implies a large scale at
which the asymptotic inner slope is reached.

\begin{figure}[tb!]
\epsscale{1.0}
\plotone{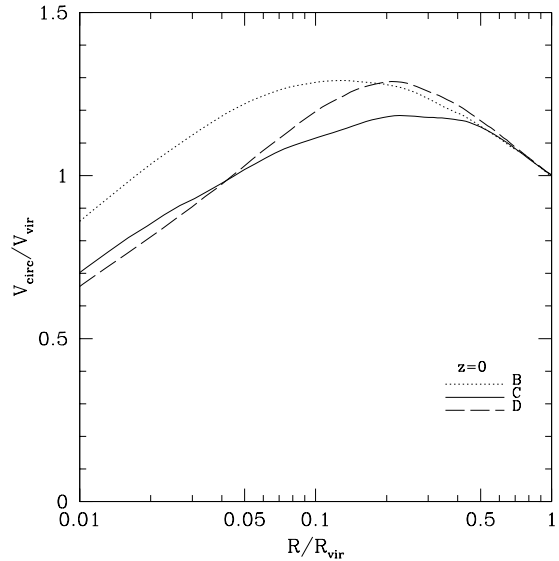}
\caption{\small Circular velocity profiles for the halos B$_1$, C$_1$,
and D$_1$
normalized to halo's virial
velocity. Halos are well resolved on all shown scales. Although the halos have
very similar masses, the profiles are very different; the differences
are due to real differences in the concentration parameters. } \label{fig:profileV}
\end{figure}

We found that substantial substructure is present inside the virial
radius in all three halos at $z=1$.  Figure~\ref{fig:Profz} shows
profiles of these halos at $z=0$ (top) and $z=1$ (bottom). The $z=0$ 
profiles are smoother than profiles at $z=1$. Note
that bumps and depressions visible in the profiles have amplitude that
is significantly larger than the shot noise.  Halo C$_3$ appeared to
be the most relaxed of the three halos.  This halo had its last major
merger somewhat earlier than the other two.  Halo D$_3$ had a major
merger event at $z\approx 2$.  A remnant of the merger is still
visible as a bump at $r\sim 100\kpch$.  The non-uniformities of
profiles caused by substructure may substantially bias analytic fits
if one uses the entire range of scales below the virial radius.
Therefore, we used only the central, presumably more relaxed, regions
in the analytic fits: $r<50\kpch$ for halo D and $r<100\kpch$ for
halos B and C (fits using only central $50\kpch$ did not change
results).

The best fit parameters were obtained by minimizing the maximum
fractional deviation of the fit: ${\rm max}[{\rm abs}(\log\rho_{\rm
fit}-\log\rho_{\rm halo})]$.  Minimizing the sum of squares of
deviations ($\chi^2$), as is often done, can result in larger errors at
small radii with the false impression that the fit fails because it has
a wrong central slope. The fit that minimizes maximum deviations
improves the NFW fit for points in the range of radii $(5-20)\kpch$,
where the NFW fit would appear to be below the data points if the fit
was done by the $\chi^2$ minimization.  For example, if we fit halo B
by minimizing $\chi^2$, the concentration slightly decreases from 12.3
(see Table~1) to 11.8, the maximum error slightly increases to 27\%,
but the fit goes below the data points for most of the points at small
radii.

We have also fitted density distribution of halo B assuming even more
stringent limits on the effects of numerical resolution.  We fitted
the halo starting at the scale equal to six times the formal
resolution, minimizing the maximum deviation.  Inside this radius
there were about 900 particles.  Resulting parameters of the fit were
close to those in Table~1: $C_{\rm NFW}=11.8$, and maximum error of
the NFW fit was 17\%.

We found that for halos B and C the errors in the Moore et al. fits
were systematically smaller than those of the NFW fits, though the
differences were not dramatic.  But Moore et al. fit poorly in the
case of halo D. It formally gave very small errors, but at the expense
of unreasonably small concentration $C_{\rm NFW}=2$. When we
constrained the approximation to have about twice larger concentration
as compared with the best NFW fit, we were able to obtain a reasonable
fit (this fit is shown in Figure~\ref{fig:Profz}). Nevertheless, the
central density distribution is fit poorly in this case.

Therefore, our analysis does not show that one analytic profile is
better then the other for description of the density distribution in
simulated halos. Despite the larger number of particles per halo and
the lower concentrations of $z=1$ halos, results are still
inconclusive.  The Moore at al.  profile is a better fit to the
profile of halo C; the NFW profile is a better fit to the central part
of the halo D. Halo B represents an intermediate case where both
profiles provide equally good fits (similar to the analysis of halo
A). Remarkably, the same conclusions hold for the halo profiles at
$z=0$.

Both at $z=0$ and $z=1$, there are real deviations in parameters of
halos of the same mass.  We find the same differences in estimates of
$C_{1/5}$ concentrations, which do not depend on specifics of an
analytic fit.  The central slope at around $1{\ }\kpc$ also changes from
halo to halo. Halos B and C have the same virial radii and
nearly the same circular velocities, yet their concentrations are
different by 30\%. Indeed, halos in the Table~3 have similar masses in
the range $(1.2-2)\times 10^{12}\Msunh$. If halos had a universal
profile -- a shape, which depends only on halo mass, then we should
expect that the circular velocity curves are very similar for our
halos.  Figure~\ref{fig:profileV} shows circular velocities for halos
B, C, and D, which have only 25\% deviations in their virial mass. The
halos clearly do not have a universal one-parameter shape. There are
substantial variations in the curves, which occur at relatively large
radii ($\sim 0.1-0.3 R_{\rm vir}$). The variations are due to
differences in halo concentration -- each curve is well described by a
NFW or Moore et al.  profile but their concentrations are somewhat
different.  Our three halos clearly constitute a small sample.
Bullock et al. (2000) and Jing (2000) studied the spread of halo concentrations in a
large sample of halos. For a given mass it was found that halos have
20-50\% variations in the concentration at $1\sigma$ level, which is
consistent with what we find for our halos.

\section{Discussion and conclusions}

We have analyzed a series of simulations with vastly different mass
and force resolutions with the goal of studying density distribution in
the central regions of galaxy-size dark matter halos. We used multiple
mass simulations performed using the ART code; the simulations were
performed with variable mass, force and temporal resolutions.  In the
highest resolution runs, we achieved a (formal) spatial dynamical
range of $2^{18}=262144$; the simulation was run with 500,000 steps
for particles at the highest level of refinement.

Using these simulations, we have studied convergence of halo density
profiles for different mass and force resolutions.  We show that the
halo profiles converge at scales larger than a certain (true numerical
resolution) scale defined by numerical effects.  This scale is
probably code dependent, but can be found for any numerical code by a
convergence study.  For the ART simulations presented here, the
density profiles converged at the scale of four times the formal force
resolution or the radius containing more than 200 particles, whichever
is larger.  In this sense, our results are consistent with results of
the ``Santa Barbara'' cluster comparison project (Frenk et al.  1999):
the density profiles of a cluster-size halo simulated with different
numerical codes and resolutions agree with each other at all
resolved scales.

In KKBP98 we have discussed convergence tests in which we varied force
resolution while keeping mass resolution fixed.  Using these tests we
concluded that density profiles converge at radii twice the local
formal resolution of a simulation.  The convergence tests presented in
this paper, however, show that mass resolution places more stringent
conditions on the trustworthy range of scales.  Although we can
reproduce our previous results (shallower than $-1$ density profiles
at radii of two formal resolution), our new convergence tests show
that these results were affected by limited mass resolution.  We
conclude that we overestimated our force resolution KKBP98 and that
the conclusions about the shallow central slopes presented there were
an artifact of that overestimate.  Other results in KKBP98 that focus
on general halo characteristics, such as the scatter in profile
shapes, and the agreement between the $V_{max}$ --- $r_{max}$
relations of simulated dark halos and dark matter dominated dwarf and
LSB galaxies, are valid.

At scales above four times the formal resolution and containing more
than 200 particles results presented in this paper agree well with
previous simulations. For example, the concentration parameter for the
halos are in good agreement with the concentration-mass dependence
($c(M)$) presented in Bullock et al.  (2000) based on the previous
simulations. 

We can also reproduce results of convergence studies by Moore et al.
(1998) and Ghigna et al. (1999). At first glance it may appear that
our conclusions are in direct contradiction with these studies that
concluded that at least several millions of particles are needed to
resolve a halo profile properly. However, the contradiction is only in
interpretation rather than in the results themselves.  For example,
Figure~2 in Ghigna et al. shows a cluster profile simulated at 3
different mass and force resolutions.  The conclusion the authors make
based on this convergence study is (at least qualitatively) similar to
our conclusion: the profiles converge at all mass resolutions at
scales above $6\epsilon$, where $\epsilon$ is the spline force
softening of their code\footnote{Note that profiles of lower
  resolution simulations perfectly agree with profile of the highest
  resolution run at scales above $3-4\epsilon$.} (roughly equivalent
to our {\em formal} resolution). Why is this criterion is more severe
than ours?  The obvious reasons are differences in the force shape and
other differences between numerical codes. However, it appears that
the softening in these simulations was set too low (the profiles were
``overresolved'') and convergence scale is determined by the number of
particles criterion rather than by force resolution. Indeed, we can
estimate the number of particles within the softening scale in these
simulations as
$N_p(<\epsilon)\approx (4\pi/3)(\rho_{crit}/m_p)(1+\delta)\epsilon^3,$
where $\epsilon$ is the softening length, $m_p$ is the particle mass,
$\rho_{crit}$ is the critical density, and $\delta$ is the overdensity
reached at the softening scales. For the HIRES simulation in Ghigna et
al., $\epsilon=0.5h^{-1}{\ }{\rm kpc}$ and $m_p=5.37\times
10^7h^{-1}{\ }{\rm M_{\odot}}$ and $\delta\sim (0.5\div 1.0)\times
  10^7$, which gives $N_p(<\epsilon)\sim 10-30$. For their LOWRES
  simulation this number is even lower $\sim 5-10$. The scale
  containing $\gtrsim 200-300$ particles should be close to the
  convergence that is found in this study (e.g., $3-4\epsilon$ for the
LOWRES run).
  
In light of these  considerations, the profiles  in Fig~2 of Moore  et
al.  (1998) are perfectly consistent with each  other if considered at
scales $\gtrsim  4\epsilon$.     Their results are   thus    perfectly
consistent with our results and  conclusions. One needs more particles
only   if the softening  is  set too small   and the inner regions are
over-resolved. In this case the  true resolution is  set by the radius
that contains at least a  couple hundred particles.  The main point is
that the higher mass  resolution is  needed to  probe deeper  into the
inner  regions of halos.   However, if one is  interested  only in the
profiles  at,  say,  $r\gtrsim  0.02R_{vir}$  (sufficient to determine
halo's concentration,    maximum  circular velocity,   etc.)  than, as
Figure~2  in Ghigna et al. (1999)  and Figure~5  in this paper clearly
show, $\sim (2-5)\times 10^5$ within the virial radius is adequate.

In this paper, we present results for halos that contain $1.2\times
10^5$ to $1.6\times 10^6$ particles within their virial radius.  We
show that for the galaxy-size halos ($M_{\rm
  vir} =7\times 10^{11}\Msunh - 2\times 10^{12}\Msunh$ and $C=9-17$)
both the NFW profile $\rho\propto r^{-1}(1+r)^{-2}$ and the Moore et
al. profile $\rho\propto r^{-1.5}(1+r^{1.5})^{-1}$ provide fairly good
fits of the simulated profiles with deviations of about 10\% for radii
larger than 1\% of the virial radius. For dwarf and LSB galaxies
commonly used for comparisons with model predictions, this corresponds
to scales $\simgreat 1$kpc.  Therefore, the debate about which
analytic profile provides a better description of the CDM halo
profiles may well be irrelevant for comparisons to measured galaxy
rotation curves.  Such comparisons are also subject to other
uncertainties, one of which is limited spatial extent of the observed
rotation curves.  The particular shape of the inner density
distribution may be important for galaxy cluster observations,
however.  Cluster-size halos are predicted to have smaller
concentrations, which means that the scale at which differences
between the NFW and Moore et al.  profiles become significant is
larger and is observationally relevant.

These results are consistent with results of Moore et al. (1999) who
found that galaxy-size halos in CDM models are well described by the
$\rho\propto r^{-1.5}(1+r^{1.5})^{-1}$ profile. The authors used
simulations of the standard CDM model with mass and force resolution
similar to our simulations ($2\times 10^6{\ }M_{\odot}$ and $1$ kpc,
respectively).  They found that the NFW profile that was fitted at
radii above $3\%$ of the halo's virial radius, underpredicts the
density at smaller radii by up to $20-30\%$. This is consistent with
out results for halos B and C, although not for halo D, which probably
indicates a certain degree of variance among profiles of halos of the
same mass (Jing 2000; Bullock et al. 2000; Avila-Reese et al. 1999).

Jing \& Suto (2000) simulated formation of galaxy-, group-, and
cluster-size halos with resolution similar to our highest-resolution
simulations. They also found that the NFW profile fit to the outer
regions (radii above few percent of the virial radius) underestimates
the density in the innermost regions of the halo. The degree of
discrepancy, however, appears to be different for different halos (see
their Fig.~2), which is consistent with our conclusions.  The authors
conclude that the shapes of the density profiles vary from galaxy- to
group-, to cluster-size halos. However, as we argued in \S~3.1, their
results could be interpreted as the manifestation of
concentration-mass relation instead.

Most recently,   Fukushige \& Makino  (2001)  simulated  12 halos with
masses  ranging from   $4.3\times 10^{11}{\  }M_{\odot}$ to $7.9\times
10^{14}{\ }M_{\odot}$, again with resolution similar to the resolution
of simulations presented here. The dependence of the inner logarithmic
slope on the halo mass  observed by Jing \& Suto  (2000) was not found
in these simulations.   Instead, the  steepest  slope of  the  density
profiles  was found to  be  close to  $-1.5$  for all masses.  This is
surprising  because   such    dependence    is    expected from     the
concentration-mass correlation (e.g., NFW; Bullock et al. 2000). It is
not clear  what could explain this  discrepancy.  Note also that setup
of these  simulations was somewhat  different from  the  setup that is
usually  used:   the  simulations   followed  halo   collapse  from  a
spherically symmetric configuration centered on a gaussian density peak
with no external tidal field included. 

We show that density profiles of halos that are not fully relaxed may
contain real non-uniformities due to substructure and differences in
the density distributions.  These non-uniformities affect the fit
quality for a particular analytic profile and result in somewhat
different values of fitted parameters (e.g., concentration). This was
clearly seen at redshift $z=1$ for halos B,C,D
(Figure~\ref{fig:Profz}), which by that time have not yet relaxed. At
redshift $z=0$ the halos were much more quiet and the deviations from
the fits were much smaller. It is interesting to note that the
non-equilibrium effects do not qualitatively change the shape of the
central parts of density profiles.  For example, we found that at both
redshifts the profile of halo C is best fit by the Moore et al.
profile. The density profile of halo D, however, is best fit by the NFW
profile again at both redshifts.  Note that these three halos were
simulated with the same mass and force resolution (indeed, in the same
simulation).  It seems that the main difference between these halos is
their merger histories.  We conclude, therefore, that differences in
merger history and/or different degree of substructure in halos of the
same mass may explain the scatter in profile shapes and concentration
parameters found in previous studies (KKBP98; Jing 2000; Bullock et
al. 2000).

In view  of the   current  less  confusing situation   regarding   the
theoretical predictions for CDM profiles, it is interesting to discuss
how the theory compares to observations.   Rotation curves of a number
of  dwarf  and   LSB galaxies have   recently   been re-examined using
H$\alpha$ observations  and/or including corrections for beam-smearing
in HI observations (e.g., Swaters,  Madore \& Trewhella 2000; van  den
Bosch et   al.   2000).  The results show   that  for the  majority of
galaxies, the H$\alpha$ rotation curves are significantly different in
their central regions than    the  rotation curves derived  from    HI
observations.  This indicates that the HI rotation curves are affected
by beam smearing  (Swaters et al. 2000).  This  also implies that beam
smearing may be at least partly responsible for the universal shape of
the LSB rotation curves discussed in KKBP98.

It is possible that part of the discrepancy between the rotation curves
of the different tracers may be due to real differences in the
kinematics of the two gas components (ionized and neutral hydrogen).
Preliminary comparisons between the new H$\alpha$ rotation curves and
model predictions show that NFW density profiles are consistent with
the observed {\em shapes} of the rotation curves (van den Bosch \&
Swaters 2000; Navarro \& Swaters 2000).  Moreover, cuspy density
profiles with inner logarithmic slopes as steep as $\sim -1.5$ also
seem to be consistent with the data (van den Bosch \& Swaters 2000).  A
separate concern is not the {\em shape} of the inner density profile,
but rather the value of the central density.  There are indications
that CDM halos are too concentrated (Navarro \& Swaters 2000; McGaugh
et al.  2000; Navarro \& Steinmetz 2000; Firmani et al. 2000) in
comparison with galactic halos.  However, van den Bosch \& Swaters
(2000) have argued, based on detailed modeling of adiabatic contraction
and beam-smearing, that dwarf galaxy concentrations are in fact
consistent with the observed distribution in {\LCDM} halos. Thus,
although the shape of galactic rotation curves may be not as different
from predictions as was thought before, the halo concentrations derived
from observations are alarmingly low. The recent observational progress
made in this field promises to resolve and clarify this issue in the
near future. At larger scales ($r\gtrsim 100$ kpc), the constraints
from weak galaxy-galaxy lensing (Fischer et al. 2000; Smith et al.
2000) should be very useful in constraining the overall profile and
concentrations of galactic halos.
 
In summary, the study presented here is aimed to clarify the issue of
convergence of the density profiles of CDM halos. We show that
convergence can be reached regardless of the mass resolution, although
the convergence scale does depend on the mass resolution: the higher
mass resolution results in smaller convergence scale for the same
objects, but it does not affect the outer parts of the profile.  Our
results also indicate that there is a real scatter in shapes of density
profiles and halo parameters.  Larger systematic studies currently
underway may put these conclusions on a firmer footing.

\acknowledgements

We acknowledge support from the grants NAG- 5-  3842 and NST- 9802787,
and also NASA and NSF grants  at UCSC.  A.V.K. acknowledges support by
NASA through Hubble Fellowship  grant  HF-01121.01-99A from the  Space
Telescope  Science Institute, which is operated  by the Association of
Universities  for  Research in Astronomy,   Inc.,  under NASA contract
NAS5-26555.  JSB was supported by  NASA LTSA  grant NAG5-3525 and  NSF
grant AST-9802568.   JRP  acknowledges  a  Humboldt  Award.   Computer
simulations presented  in this paper were done  at the National Center
for Supercomputing Applications (NCSA), Urbana-Champaign, Illinois and
at the Leibniz-Rechenzentrum in Munich.

%===================================================================

\end{document}